\documentclass[preprint,showpacs,preprintnumbers,amsmath,amssymb]{revtex4}
\usepackage{bm}
\usepackage{epsf}

\everymath{\displaystyle}
\begin{document}           
\title{Validation of the Eriksen Method
of the Exact Foldy–Wouthuysen Representation}

\author{A.J. Silenko}
\affiliation{Research Institute for Nuclear Problems, Belarusian
State University, Minsk, 220030 Belarus}

\begin{abstract}
The Eriksen method is proven to yield a correct and exact result
when a sufficient condition of exact transformation to the
Foldy-Wouthuysen (FW) representation is satisfied. Therefore, the
Eriksen method is confirmed as valid. This makes it possible to
establish the limits within which the approximate "step-by-step"
methods are applicable. The latter is done by comparing the
relativistic formulas for a Hamiltonian operator in the FW
representation (obtained using those methods) and the known
expression for the first terms of a series, which defines the
expansion of this operator in powers of $v/c$ as found by applying
the Eriksen method.
\end{abstract}

\pacs {03.65.Pm, 11.10.Ef} \maketitle

The transition to the FW representation (FW transformation)
carried out in well-known paper \cite{FW} is a unitary
transformation leading to an even Hamiltonian operator that
characterizes the FW representation. A Hamiltonian operator is
block-diagonal in this representation; i.e., it is diagonal in two
spinors or their analogues for particles with spins $S\neq1/2$.
The FW representation has unique properties that render it special
in quantum mechanics. In this representation the quantum
mechanical operators for relativistic particles in an external
field have the same form as in the nonrelativistic quantum theory.
The relationships between operators are similar to those between
respective classical quantities. The specified properties of the
FW representation make it impossible for ambiguities to emerge
when this representation is used to go over to the quasiclassical
approximation and classical limit of relativistic quantum
mechanics \cite{FW,CMcK}.

Nonetheless, it was noticed a relatively long time ago that the
block diagonalization of the Hamiltonian is not at all identical
to the transition to FW representation (see Ref. \cite{JMPcond}
and references therein). Moreover, strictly speaking, even the
method developed by Foldy and Wouthuysen \cite{FW} does not lead
to this representation \cite{EK}. It is a step-by-step method.
When using this kind of method, a transition to the block-diagonal
form of the Hamiltonian is achieved by successive iterations,
resulting in the elimination of odd (nondiagonal) highest order
terms at each step. However, the operator of the exact FW
transformation $ U_{FW} ~(\Psi_{FW}=U_{FW}\Psi)$ for spin-1/2
particles must satisfy the Eriksen condition \cite{E}:
\begin{eqnarray}
\beta U_{FW}=U^\dag_{FW}\beta, \label{pdnvEri}
\end{eqnarray}
where $\beta$ is a Dirac matrix. When the operator $U_{FW}$ is
represented in the exponential form
\begin{eqnarray}
U_{FW}=\exp{(iS)} \label{expfEri}
\end{eqnarray}
condition (\ref{pdnvEri}) is equivalent to requiring that the
exponential operator $S$ is Hermitian and odd \cite{EK}. By virtue
of the Hausdorff theorem \cite{Hausdorff}, the step-by-step
methods are shown \cite{EK} to not satisfy the $S$ operator
oddness condition. Therefore, they can ensure only an approximate
transition to the FW representation.

An original Hamiltonian for spin-1/2 particles can be represented
in the following general form:
\begin{equation} {\cal H}_D=\beta m+{\cal E}+{\cal
O},~~~\beta{\cal E}={\cal E}\beta, ~~~\beta{\cal O}=-{\cal
O}\beta, \label{2eq3} \end{equation} where ${\cal E}$ and ${\cal
O}$ are even and odd operators, respectively.

The general form of the operator of exact FW representation was
found by Eriksen \cite{E}:
\begin{equation}
U_{FW}=\frac12(1+\beta\lambda)\left[1+\frac14(\beta\lambda+\lambda\beta-2)\right]^{-1/2},
~~~ \lambda=\frac{{\cal H}_D}{\sqrt{{\cal H}_D^2}}, \label{eq3EI}
\end{equation}
where $\lambda=+1$ and $-1$ for solutions with positive and
negative energies, respectively. It is important that \cite{E}
\begin{equation}\lambda^2=1, ~~~ [\beta\lambda,\lambda\beta]=0 \label{eq3.12}
\end{equation}
and the operator $\beta\lambda+\lambda\beta$ is even:
\begin{equation}[\beta,(\beta\lambda+\lambda\beta)]=0.\label{eqEII}
\end{equation}
The even operators are block diagonal and do not mix upper and
lower spinors. Formula (\ref{eq3EI}) can also be represented as
follows \cite{JMPcond}:
\begin{equation}
U_{FW}=\frac{1+\beta\lambda}{\sqrt{(1+\beta\lambda)^\dag(1+\beta\lambda)}}.
\label{eq3.15}
\end{equation}
Two operator factors in the denominator commute.

The validity of the Eriksen transformation was argued in
\cite{VJ}. The operator $U_{FW}$ turns to zero either at lower or
upper spinors of any eigenfunction of the Dirac Hamiltonian. This
transformation is carried out in just a single step. However, it
is problematic to effectively use the Eriksen method to find
relativistic formulas for particles in an external field, since
the general formula (\ref{eq3EI}) is rather cumbersome and
contains square roots of Dirac matrices. An expression for the
Hamiltonian operator in FW representation was found \cite{VJ} only
in the form of a relativistic correction series in powers of
${\cal E}/m,~{\cal O}/m$. Thus, the Eriksen method does not solve
the problem of finding compact relativistic expressions for the
Hamiltonian operator in this representation.

Some of the step-by-step methods give those relativistic
expressions \cite{B,TMP,JMP,PRA,Gos,Gos2,GosHM,GosM}. Note that
the method developed in \cite{PRA} is applicable to the case of
particles with arbitrary spin. Since all the methods mentioned so
far are approximate, it is necessary to determine the limits of
their applicability. It is clear that the simplest and most
reliable way to make such a determination is to compare
relativistic Hamiltonians in the FW representation obtained by
using the step-by-step methods with exact expansion into a series
presented in \cite{VJ}.

However, it is first needed to make a maximally possible
verification of the Eriksen (\ref{eq3EI}) or equivalent
(\ref{eq3.15}) formulas because a justification of the oddness of
the exponent operator $S$, which is part of formula
(\ref{expfEri}) and is given in \cite{EK}, is not rigorous
mathematical proof. The fact that the Eriksen formula gives a
correct answer in the case of free particles \cite{E} is also not
enough to prove its validity.

Back in 2003, a sufficient condition for carrying out an exact FW
transformation was found in Ref. \cite{JMP}. This transformation
is exact if an external field is stationary and the operators
${\cal E}$ and ${\cal O}$ commute:
\begin{equation}
[{\cal E},{\cal O}]=0.
\label{2eq14}
\end{equation}
In this case
\begin{equation}
{\cal H}_{FW}=\beta \epsilon+{\cal E},~~~\epsilon=\sqrt{m^2+{\cal O}^2}.
\label{2eq17}
\end{equation}

In the present paper we find out if the Eriksen formula is
compatible with this exact transformation.

It should be indicated that, when finding a square root of
operators, an obvious condition ${(\sqrt{A})}^2=A$ should be
supplemented by the condition of equality between a square root of
unit matrix and the unit matrix itself \cite{JMP}. Thus, for free
particles ${\cal E}=0,~{\cal O}=\bm\alpha\cdot\bm p$ and
$\lambda=(\beta m+ \alpha\cdot\bm p)/\sqrt{m^2+\bm p^2}$.

With this in mind and considering that the operators ${\cal E}$
and ${\cal O}$ commute, a square root can be brought to the form
\begin{equation}
\sqrt{{\cal H}_D^2}=\epsilon+\frac{(\beta m+{\cal O}){\cal E}}{\epsilon}=
\epsilon\left[1+\frac{(\beta m+{\cal O}){\cal E}}{\epsilon^2}\right].
\label{qroot}
\end{equation}

Since the following equality
$$ {\cal H}_D=(\beta m+{\cal O})\left[1+\frac{(\beta m+{\cal O}){\cal E}}{\epsilon^2}\right],
$$
is correct, the sign operator $\lambda$ assumes the form
\begin{equation}
\lambda=\frac{\beta m+{\cal O}}{\epsilon}.
\label{znak}
\end{equation}
It is very important that in the case at hand it is independent of
${\cal E}$.

The operator of exact FW transformation is given by
\begin{equation}
U_{FW}=\frac{\epsilon+m+\beta{\cal
O}}{\sqrt{2\epsilon(\epsilon+m)}}. \label{eq18} \end{equation}
Formula (\ref{eq18}) has the same form as a respective one found
in \cite{JMP}. It is natural that the expression for the
Hamiltonian operator in FW representation obtained using the
Eriksen method coincides with the result deduced in \cite{JMP}
provided that condition (\ref{2eq14}) is satisfied:
\begin{equation}
{\cal H}_{FW}=\beta \epsilon+{\cal E}.
\label{eq17}
\end{equation}

Thus, in the case under consideration, the Eriksen method gives a
correct and exact result. It is natural that, under the additional
condition (\ref{2eq3}), the original Dirac Hamiltonian
(\ref{2eq14}) has a much more general form than that for free
particles and describes a number of practically important cases
(see Refs. \cite{PRA,PRD2}). Therefore, this confirmation of the
validity of the Eriksen method considerably improves the situation
with its justification.

The transformation considered above, which is equivalent to taking
a square root in the operator sense, can be performed in a general
case. If the operators ${\cal E}$ and ${\cal O}$ do not commute,
then, in the weak field approximation $(|{\cal E}|\ll m)$ and
taking into account only unary and binary commutators,
\begin{equation}
\sqrt{{\cal
H}_D^2}=\epsilon+\frac14\left\{\frac{1}{\epsilon},\left\{(\beta
m+{\cal O}),{\cal E}\right \}\right \}-\frac18\left\{\frac{\beta
m+{\cal O}}{\epsilon},\left[\epsilon ,[\epsilon , {\cal E}]\right
]\right \}. \label{rootg}
\end{equation}

Even in this case the subsequent calculations become rather
cumbersome, although they can be carried out analytically for
certain specific problems. In the case of abandoning the weak
field approximation, the expression for $\sqrt{{\cal H}_D^2}$
itself becomes rather cumbersome.

One important advantage of step-by-step methods is that they
require far fewer calculations to be done. As a consequence, these
methods were successfully used to carry out the FW transformation
in describing the interaction of relativistic particles with
external fields (see Refs. \cite{JMP,OSTRONG} and references
therein). Finding a correct and exact expression for the
Hamiltonian operator in the present paper by using the Eriksen
method and under condition (\ref{2eq14}) gives good reason to
consider that the results derived in \cite{VJ} are exact ones and,
therefore, makes it possible to set the limits of applicability of
step-by-step methods. This is done by comparing the relativistic
formulas for a Hamiltonian operator in FW representation, which is
obtained by using step-by-step methods with the known expression
for the first terms of a series defining the expansion of this
operator in powers of $v/c$ found in \cite{VJ} by applying the
Eriksen method.

\medskip\textbf{Acknowledgments.} This work was supported by
the Belarusian Republican Foundation for Fundamental Research,
grant No. $\Phi$12D-002.

\end{document}